\documentclass[proof]{WileyASNA-v1}

\articletype{Proceedings}%

\received{XX month YYYY}
\revised{XX month YYYY}
\accepted{XX month YYYY}

\usepackage[x11names]{xcolor}
    
\usepackage{lipsum}
\usepackage{comment}

\usepackage{xfrac}
\usepackage{mathtools}
\usepackage{dsfont}

\usepackage{pgfplots}
  \pgfplotsset{compat=newest}



\usepackage{float}

\usepackage{upgreek}
\usepackage{bbm}

\usepackage{pgfplotstable}
      \usepgfplotslibrary{fillbetween}

\begin{document}

\title{Neutrino oscillation in a minimal length spacetime}

\author[1]{Luísa Nadal Camargo\,*}
\author[2a,2b]{Thiago Oliveira Ferreira}
\author[1]{Dimiter Hadjimichef}
\author[1]{César A. Z. Vasconcellos}
\authormark{Nadal-Camargo et al.}

\address[1]{\orgdiv{Instituto de Física}, \orgname{Universidade Federal do Rio Grande do Sul (UFRGS)}, \orgaddress{\state{Porto Alegre}, \country{Brazil}}}
\address[2a]{\orgdiv{Department of Applied Mathematics}, \orgname{University of Waterloo}, \country{Canada}}
\address[2b]{\orgdiv{Perimeter Institute for Theoretical Physics}, \country{Canada}}

\corres{*\email{luisa.nadal.c@gmail.com}}


\abstract{

We investigate how neutrino oscillations are modified in a non-commutative spacetime characterized by a minimal length scale, described by the Quesne–Tkachuk algebra. By incorporating the algebra’s deformation parameter $\beta$ into the effective neutrino mass, we derive the 2-flavor oscillation probability in this non-commutative setting. The resulting probability depends not only on the usual mass-squared difference but also on a fourth-order mass difference scaled by $\beta$. A comparison between the standard and non-commutative oscillation probabilities reveals a beat pattern arising from the additional non-commutative phase, which induces a small shift in the oscillation profiles. Finally, we extend our analysis to include the effects of a magnetic field on neutrino propagation.

}
 

\maketitle



\section{Introduction}

The first hint on the neutrino oscillation phenomenon came from the abnormal flux of solar neutrinos: although the Sun produces only electron neutrinos, measurements from early detectors, such as Homestake (1960s to 1980s), had a deficit between half and two-thirds the predicted flux \citep{DAVIS199413}. This suggested either flawed solar models or new particle physics.

In 2001, the Sudbury Neutrino Observatory (SNO) resolved the issue by detecting the total neutrino flux, confirming that approximately $2/3$ of solar electron neutrinos oscillate into muon and tau neutrinos during their journey to Earth \citep{Poon_2002}. Nowadays, such phenomenon is called \textit{neutrino oscillation} and it implies that neutrinos have non-zero mass, requiring an extension of the Standard Model.

The hypothesis of massive neutrinos suggests non-trivial electromagnetic properties. Recent experiments, including the Borexino collaboration, constrain the neutrino magnetic moment to $\lesssim 10^{-11}\mu_\text{B}$, where $\mu_\text{B}$ is the Bohr magneton \citep{Borexino_2017}. While this is undetectably small in terrestrial magnetic fields, its value becomes significant in extreme environments, where $ \boldsymbol{B} \sim 10^{11}$ T, typical of pulsars, where neutrino spin can precess. Thus, in addition to flavor oscillation, neutrinos could also present spin oscillation under a transverse magnetic field, i.e. a left-handed neutrino may convert into a right-handed neutrino. 

In parallel to that, in the early days of quantum field theory and before the dawn of renormalization, Werner Heisenberg proposed that, at extremely small length scales, spacetime coordinates could be treated as non-commutative operators. This approach would naturally introduce an effective UV cutoff, potentially resolving the problematic divergences that afflicted quantum field theories. This idea was first formalized by \cite{Snyder:1946qz}, and became particularly relevant when brought up by Kempf in the late 1990s \citep{Kempf:1994, Kempf_1995}. Nevertheless, there is an undesirable feature in the Kempf algebra: it is not Lorentz-covariant. This limitation was addressed by \cite{quesne_tkachuk_2006}, who extended Kempf’s algebra.

Our goal in this paper is to study the neutrino oscillation, both in vacuum and in a constant background magnetic field, in the context of a minimal length spacetime, described by the Quesne--Tkachuk algebra, as initiated in \cite{Thiago-Ferreira:2023}. We reach this goal in the following way: in Section \ref{sec2}, we review the Quesne--Tkachuk algebra for describing the minimal length spacetime, the usual neutrino oscillation in vacuum and the modified Dirac equation and its solutions, as first found by \cite{moayedi-2011}. In Section \ref{sec3}, we analyze how the oscillation of neutrinos is affected by the presence of a minimal length. In Section \ref{sec4}, we extend our computations in the presence of a background magnetic field, inspired by \cite{Popov_2019}. Finally, in Section \ref{sec5}, we present our conclusions. Einstein summation is assumed everywhere, and we set $c=1$.


\section{Preliminaries}\label{sec2}
In this section, we review the necessary background from the literature. These results will be used to introduce minimal length spacetime effects into neutrino oscillations.

\subsection{Minimal Length Spacetime}

The Quesne--Tkachuk algebra for a $(D+1)$-dimensional spacetime is described by the following commutation relations:
\begin{subequations} \label{eq:2-Quesne-Tkachuk-algebra}
\begin{equation}\label{eq:2-Quesne-Tkachuk-algebra-commutators-XP}
    [\hat{X}^\mu, \hat{P}^\nu] = -\mathrm{i}\hbar\bigl((1-\beta\hat{P}_\rho \hat{P}^\rho)g^{\mu\nu} - \beta'\hat{P}^\mu\hat{P}^\nu\bigr),
\end{equation}
\begin{equation}\label{eq:2-Quesne-Tkachuk-algebra-commutators-XX}
    \begin{aligned}
    [\hat{X}^\mu, \hat{X}^\nu] = \mathrm{i}\hbar\frac{2\beta-\beta' -(2\beta+\beta')\beta\hat{P}_{\rho}\hat{P}^\rho}{1-\beta\hat{P}_{\rho}\hat{P}^\rho}\bigl(\hat{P}^\mu\hat{X}^\nu 
    - \hat{P}^\nu\hat{X}^\mu\bigr),
    \end{aligned}
\end{equation}
\begin{equation}\label{eq:2-Quesne-Tkachuk-algebra-commutators-PP}
    [\hat{P}^\mu, \hat{P}^\nu] = 0,
\end{equation}
\end{subequations}
where $\beta$ and $\beta'$ are very small positive parameters that represent spacetime deformation, and $g^{\mu\nu} = \text{diag}\{+1, -1, ..., -1\}$ is the Minkowski metric \citep{quesne_tkachuk_2006}.

The algebra \eqref{eq:2-Quesne-Tkachuk-algebra} is Lorentz-covariant; it is enough, then, to calculate the generalized uncertainty principle for the \textit{i}-th component of position and momentum, for $i \in \{1,2,...,D\}$, by taking the expected value of (\ref{eq:2-Quesne-Tkachuk-algebra-commutators-XP}):
\begin{equation*}
    \bigl\langle [\hat{X}^i, \hat{P}^i] \bigr\rangle = \mathrm{i}\hbar \bigl(1 - \beta \langle \hat{P}_\rho \hat{P}^\rho \rangle + \beta'\langle \hat{P}^i\hat{P}^i \rangle\bigr),
\end{equation*}
which leads to
\begin{equation*}
    \begin{aligned}
         \Delta \hat{X}^i \Delta \hat{P}^i &\geq \frac{\hbar^2}{2}\Biggl|1 - \beta\biggl[\langle (\hat{P}^0)^2 - 
         \sum_{j=1}^D\bigl[(\Delta \hat{P}^j)^2 + \langle \hat{P}^j\rangle^2 \bigr] \biggr]+ {}\\
         &\qquad+\beta'\bigl[(\Delta \hat{P}^i)^2 + \langle \hat{P}^i\rangle^2 \bigr] \Biggr|.
     \end{aligned}
\end{equation*}

When assuming $\Delta \hat{P}^j = \Delta \hat{P}$ for $j\in\{1,2,...,D\}$ (i.e. assuming isotropic uncertainties), one can isolate $\Delta \hat{X}^i$; extremizing this expression, one finds that the minimal value in position is
\begin{equation*}
    \begin{aligned}
    \Delta \hat{X}^i_\text{min} &= \hbar\sqrt{\Biggl[1-\beta\biggl(\langle (\hat{P}^0)^2\rangle - \sum_{j=1}^D\langle P^j\rangle^2 \biggr) + \beta'\langle P^i\rangle^2 \Biggr]} \times {}
    \\ &\qquad\times {}
    \sqrt{(D\beta+\beta')} .
    \end{aligned}
\end{equation*}
The isotropic absolutely smallest uncertainty in position, the \textit{minimal length}, is found when assuming $\langle \hat{P}^j\rangle = 0$:
\begin{equation}\label{eq:2-Q-T-minimal-length}
    \ell_{\text{min}}
        = (\Delta X)_0 = \hbar \sqrt{(D\beta + \beta') \bigl[1 - \beta\langle (\hat{P}^0)^2\rangle\bigr]}.
\end{equation}

For $D=3$ spatial dimensions and $\beta' = 2\beta$, scenario in which the usual position commutator in \eqref{eq:2-Quesne-Tkachuk-algebra-commutators-XX} is recovered, the Quesne--Tkachuk algebra reduces, up to order of $\beta$, to
\begin{subequations} \label{eq:2-Quesne-Tkachuk-algebra-beta'=2beta}
\begin{equation}\label{eq:2-Quesne-Tkachuk-algebra-commutators-XP-beta'=2beta}
    [\hat{X}^\mu, \hat{P}^\nu] = -\mathrm{i}\hbar\bigl(\bigl(1-\beta\hat{P}_\rho \hat{P}^\rho\bigr)g^{\mu\nu} -2 \beta\hat{P}^\mu\hat{P}^\nu\bigr),
\end{equation}
\begin{equation}\label{eq:2-Quesne-Tkachuk-algebra-commutators-XX-beta'=2beta}
    [\hat{X}^\mu, \hat{X}^\nu] = 0,
\end{equation}
\begin{equation}\label{eq:2-Quesne-Tkachuk-algebra-commutators-PP-beta'=2beta}
    [\hat{P}^\mu, \hat{P}^\nu] = 0.
\end{equation}
\end{subequations}

\cite{Samar-tkachuk:2010sxi} found a representation for $\hat{X}^\mu$ and $\hat{P}^\mu$ that satisfies the commutation relations in equations \eqref{eq:2-Quesne-Tkachuk-algebra-beta'=2beta} keeping the position operators unchanged, but correcting the momentum operators due to the presence of the minimal length. The Samar--Tkachuk representation reads as follows:
\begin{subequations}
\begin{equation}\label{eq:2-X-operator-samar-tkachuk-representation-beta'=2beta}
    \hat{X}^\mu = x^\mu,
\end{equation}
\begin{equation}\label{eq:2-P-operator-samar-tkachuk-representation-beta'=2beta}
    \hat{P}^\mu = \bigl(1-\beta p_\nu p^\nu\bigr)p^\mu,
\end{equation}
\end{subequations}
where $x^\mu$ and $p^\mu$ are the ordinary position and momentum operators, that follow $[x^\mu, p^\mu] = -\mathrm{i}\hbar g^{\mu\nu}$, $[x^\mu, x^\nu] = [p^\mu, p^\nu] = 0$, and $\hat{X}^\mu$ and $\hat{P}^\mu$ are ($3+1$)-dimensional vectors, for $D=3$.



\subsection{Neutrino oscillation in vacuum}

A neutrino flavor eigenstate $|\upnu_\alpha \rangle$ is described as a superposition of mass eigenstates $|\upnu_j\rangle$, weighted by elements of the Pontecorvo--Maki--Nakagawa--Sakata (PMNS) matrix,
\begin{equation}\label{eq:3-flavor-as-mass-mixing}
    |\upnu_\alpha\rangle = \sum_j U_{\alpha j}^* |\upnu_j\rangle.
\end{equation}
$U$ is a unitary mixing matrix, which ensures orthonormality of flavor eigenstates. 

Massive neutrinos are eigenstates of the Hamiltonian,
\begin{equation*}
    \hat{H}|\upnu_j\rangle = E_j |\upnu_j\rangle,
\end{equation*}
where the energy eigenvalues are given by the usual dispersion relation
\begin{equation}\label{eq:3-energy-dispersion-relation}
    E_j = \sqrt{\boldsymbol{p}^2 + (m_j)^2}.
\end{equation}

The Hamiltonian guarantees that the massive neutrino states evolve in time as plane waves
\begin{equation}\label{eq:3-time-evolution-neutrino-mass-state}
    |\upnu_j(t)\rangle = \mathrm{e}^{-\mathrm{i}E_jt/\hbar} |\upnu_j\rangle,
\end{equation}
where $|\upnu_j (t=0)\rangle = |\upnu_j\rangle$. Substituting \eqref{eq:3-time-evolution-neutrino-mass-state} into \eqref{eq:3-flavor-as-mass-mixing}, we obtain
\begin{equation}\label{eq:3-time-evolution-mixing-flavor-mass-before}
    |\upnu_\alpha(t)\rangle = \sum_j U^*_{\alpha j}\,\mathrm{e}^{-\mathrm{i}E_jt/\hbar}|\upnu_j\rangle.
\end{equation}

The amplitude of the flavor transition $\upnu_\alpha \to \upnu_\beta$ as a function of time is given by
\begin{equation}\label{eq:3-amplitude-of-transition}
    \langle \upnu_\beta | \upnu_\alpha(t)\rangle = \sum_j U^*_{\alpha j} U_{\beta j}^{} \,\mathrm{e}^{-\mathrm{i}E_jt/\hbar} ,
\end{equation}
so that the transition probability is
\begin{equation*}
    \begin{aligned}
        \mathcal{P}_{\alpha \to \beta}(t) \equiv \bigl|\langle \upnu_\beta | \upnu_\alpha(t)\rangle\bigr|^2 =
         \sum_{j,k} U^*_{\alpha j} U_{\beta j}^{} U_{\alpha k}^{} U^*_{\beta k}\,\mathrm{e}^{-\mathrm{i}(E_j-E_k)t/\hbar}.
    \end{aligned}
\end{equation*}

The neutrino ultrarelativistic scenario ensures that the dispersion relation \eqref{eq:3-energy-dispersion-relation} can be appropriately approximated as 
\begin{equation*}
    E_j \approx E + \frac{(m_j)^2}{2E},
\end{equation*}
where $E = |\boldsymbol{p}|$. So,
\begin{equation*}
    E_j - E_k \approx \frac{\Delta m^2_{jk}}{2E}, \qquad \Delta m_{jk}^2 = m_j^2 - m_k^2.
\end{equation*}
$\Delta m_{jk}^2$ is the mass-squared difference. In addition to that, ultrarelativistic neutrinos propagate almost with the speed of light, so it is possible to approximate the traveling time $t$ with the distance traveled, $t\approx L$. Thus, the neutrino oscillation probability becomes
\begin{equation}\label{eq:3-3-neutrino-oscillation-probability}
    \begin{aligned}
     \mathcal{P}_{\alpha \to \beta}(L, E)
     =\sum_{j,k} U^*_{\alpha j} U_{\beta j}^{} U_{\alpha k}^{} U^*_{\beta k}\exp\biggl({-\mathrm{i}\frac{\Delta m_{jk}^2}{2E}\frac{L}{\hbar}}\biggr).
     \end{aligned}
\end{equation}

As commonly found in the literature, we restrict ourselves the simplifying toy model of two-flavor neutrino mixing,
i.e., we consider only two massive neutrino states and two flavor neutrinos --- namely, the electron ($\mathsf{e}$) and the muon ($\upmu$) neutrinos. In this scenario, the mixing matrix simplifies to
\begin{equation}\label{eq:3-2-times-2-mixing-matrix}
    U = \begin{pmatrix}
        \cos\theta & \sin\theta \\
        -\sin\theta & \cos\theta
    \end{pmatrix} = U^*.
\end{equation}
Hence, the probability of the transition $\upnu_\mathrm{e} \to \upnu_\upmu$ for the two-flavor neutrino model is
\begin{equation}\label{eq:3-oscillation-probability-1-cos}
    \begin{aligned}
    \mathcal{P}_{\alpha\to\beta}(L,E)= 
    \frac{1}{2}\sin^2(2\theta)\biggl[1-\cos\biggl(\frac{\Delta m^2}{2E}\frac{L}{\hbar}\biggr) \biggr],
    \end{aligned}
\end{equation}
where $\Delta m^2 \equiv m_2^2 - m_1^2$, and $m_2^2 > m_1^2$.
    

\subsection{Modified Dirac Equation}\label{subsec-modified-dirac}

To apply the minimal length spacetime to the usual neutrino oscillation in vacuum, we follow the approach in \cite{moayedi-2011}. We begin by considering the Dirac Lagrangian density for a spinor field $\Psi$ with spin-$\frac{1}{2}$,
\begin{equation}\label{eq:4-expanded-lagrangian-density}
    \mathcal{L}_{\text{Dirac}}
        = \frac{\mathrm{i}\hbar}{2} \Bigl[ \bar{\Psi}\,\gamma^\mu(\partial_\mu \Psi) - (\partial_\mu \bar{\Psi})\,\gamma^\mu\,\Psi \Bigr] -m\,\bar{\Psi}\Psi,
\end{equation}
where $\bar{\Psi}$ is its corresponding adjoint spinor. In the Samar--Tkachuk representation for first order in $\beta$, we have
\begin{subequations} \label{eq:4-Samar-Tkachuk}
\begin{equation}\label{eq:4-x-samar-tkachuk}
    x^\mu \mapsto x^\mu,
\end{equation}
\begin{equation}\label{eq:4-partial-mu-samar-tkachuk}
    \partial^\mu \mapsto \bigl(1+\beta\hbar^2\square\bigr)\,\partial^\mu,
\end{equation}
\end{subequations}
where $\square \equiv g^{\mu\nu}\partial_\nu\partial_\mu$ is the d'Alambertian operator. Substituting \eqref{eq:4-Samar-Tkachuk} into the Lagrangian density in (\ref{eq:4-expanded-lagrangian-density}), it follows that
\begin{equation}\label{eq:4-lagrangian-density-modified}
    \begin{aligned}
    \mathcal{L}_{\text{Dirac}}^{\mathrm{NC}}
        &= \mathcal{L}_{\text{Dirac}}^{}
            + \frac{\mathrm{i}\hbar^3}{2}\beta \Bigl[ \bar{\Psi}\,\gamma^\mu \bigl(\square\,\partial_\mu \Psi\bigr) - \bigl(\square\,\partial_\mu \bar{\Psi}\bigr)\,\gamma^\mu\,\Psi \Bigr].
    \end{aligned}
\end{equation}
The new term indicates the correction of the minimal length in \eqref{eq:4-expanded-lagrangian-density}. The modified Dirac equation is the equation of motion associated with \eqref{eq:4-lagrangian-density-modified}.

The \textit{modified Dirac equation} is the equation of motion associated to \eqref{eq:4-lagrangian-density-modified}, which reads as
\begin{equation}\label{eq:4-modified-dirac-equation}
    \bigl[ \mathrm{i}\hbar\gamma^\mu \bigl( 1 + \beta\hbar^2\square \bigr) \partial_\mu - m\mathbbm{1} \bigr] \Psi = 0 .
\end{equation}
The term $\mathrm{i}\hbar^3\beta\,\gamma^\mu\square\,\partial_\mu\Psi$ indicates the minimal length effects. 


To obtain the plane wave solution for the modified Dirac equation \eqref{eq:4-modified-dirac-equation}, consider the ansatz
\begin{equation}\label{eq:4-plane-wave-ansatz}
    \Psi =\psi \exp\biggl(-\frac{\mathrm{i}}{\hbar} \, p_\mu x^\mu \biggr) = \psi\exp\biggl[\frac{\mathrm{i}}{\hbar}\bigl(\boldsymbol{p}\cdot\boldsymbol{x}-Et\bigr) \biggr],
\end{equation}
where $\psi$ is a constant spinor.

Substituting the ansatz \eqref{eq:4-plane-wave-ansatz} into \eqref{eq:4-modified-dirac-equation} and using the properties of the Pauli--Dirac matrices, we find
\begin{equation}\label{eq:4-modified-dirac-plane-wave-ansatz}
     \biggl\{ \Bigl[ 1- \beta \bigl(E^2 - \boldsymbol{p}^2\bigr) \Bigr] \bigl(E\mathbbm{1} - \hat{\boldsymbol{\alpha}}\cdot\boldsymbol{p} \bigr) -m \gamma^0 \biggr\}\,\psi =0.
\end{equation}
For non-trivial $\psi$, the determinant of the matrix in curly brackets must vanish, which happens when
\begin{equation}\label{eq:4-modified-energy-momentum-equation}
    \Bigl[ 1 - \beta \bigl( E^2 - \boldsymbol{p}^2 \bigr) \Bigr]^2 \bigl(E^2 - \boldsymbol{p}^2\bigr) - m^2 = 0.
\end{equation}
Equation (\ref{eq:4-modified-energy-momentum-equation}) is the \textit{modified energy-momentum relation}. Note that when $\beta =0$ we obtain the usual energy-momentum relation \eqref{eq:3-energy-dispersion-relation}. 

Let us now analyse the scenario wherein $\beta$ is non-zero, for first order in $\beta$. Define the \textit{effective mass} $\tilde{m}$ as the one satisfying the usual dispersion relation \eqref{eq:3-energy-dispersion-relation}, i.e.
\begin{equation}\label{eq:4-effective-mass}
    E^2 = \boldsymbol{p}^2 + \tilde{m}^2
    \ \Longrightarrow \
    \tilde{m} \equiv \sqrt{E^2 - \boldsymbol{p}^2} .
\end{equation}
Substituting $E^2$ from (\ref{eq:4-effective-mass}) into (\ref{eq:4-modified-energy-momentum-equation}), ignoring $\mathcal{O}(\beta^2)$ and the negative values, it is possible to write the effective mass of a fermion in a minimal length space time as:
\begin{equation}\label{eq:4-effective-mass-moayed}
    \tilde{m}_{\pm} = \sqrt{\frac{1}{4\beta} \Bigl( 1\pm\sqrt{1-8\beta m^2} \Bigr)},
\end{equation}
where $m$ in the right-hand size represents the \textit{kinetic mass} in the Lagrangian density \eqref{eq:4-lagrangian-density-modified}.

For small values of $\beta$, we have
\begin{equation}\label{eq:4-mu-minus-expanded}
    \tilde{m}_- = m + \beta m^3 + \mathcal{O}(\beta^2).
\end{equation}
The expansion for $\tilde{m}_+$ includes terms of order $\beta^{-1/2}$, which would lead to an excessively large effective mass. Since this work focuses on standard, light neutrinos, the $\tilde{m}_+$ solution is discarded. Contributions from this solution are excluded because they pertain to energy scales far beyond current experimental neutrino energies and potentially violate self-consistency of the theory, as discussed in \cite{Thiago-Ferreira:2023}. Hence, this scenario is not relevant to our analysis.


\section{Vacuum Neutrino Oscillation in a Minimal Length Spacetime}\label{sec3}
Now, let us apply solution found by \cite{moayedi-2011} into the usual neutrino oscillation, described in Subsection \ref{subsec-modified-dirac}. Recall that the standard two-flavor oscillation probability is given by
\begin{equation}\label{eq:4-probability-oscillation-commutative}
     \mathcal{P}_{\mathsf{e} \to \upmu}^{\text{C}}(L, E) = \sin^2(2\theta)\sin^2(W_\text{C}),
\end{equation}
where the conventional oscillation phase is
\begin{equation}
    W_\text{C} = \frac{\Delta m^2}{4\hbar}\frac{L}{E}.
\end{equation} 

To include non-commutative spacetime effects in neutrino oscillations, we replace the standard masses $m_i$ in the mass-squared difference $\Delta m^2 \equiv m_i^2 - m_j^2$ by the effective masses $\tilde{m}_i^{(-)}$ from the previous section. Up to first order in $\beta$, this yields
\begin{equation*}
    \begin{aligned}
    \Delta m^2
        \mapsto (\tilde{m}_i^-)^2 - (\tilde{m}_j^-)^2
        &= \bigl( m_i + \beta m_i^3 \bigr)^2
            - \bigl( m_j + \beta m_j^3 \bigr)^2 \\
        & = \Delta m^2 + 2\beta \Delta m^4,
    \end{aligned}
\end{equation*}
where $\Delta m^4 \equiv m_i^4 - m_j^4$. This modifies the oscillation phase to:
\begin{equation*}
    W_\text{NC}
        = \frac{1}{4\hbar} \frac{L}{E} \bigl(\Delta m^2 + 2\beta \Delta m^4\bigr)
        = W_\text{C} + \delta_\text{NC}.
\end{equation*}
Thus, the non-commutative phase $W_\text{NC}$ is given by the standard phase $W_\text{C}$ plus a $\beta$-proportional correction $\delta_\text{NC}$, which encodes the effects of a minimal length scale in spacetime.

Hence, a neutrino that travels through the non-commutative spacetime has probability of oscillation given by
\begin{align}
    \mathcal{P}_{\mathsf{e} \to \upmu}^{\text{NC}}(L, E)
        & = \sin^2(2\theta)\sin^2\bigl(W_\text{C} + \delta_\text{NC}\bigr)
    \nonumber \\
        &=\sin^2(2\theta)\sin^2\biggl[\frac{1}{4\hbar}\frac{L}{E} \bigl(\Delta m^2 + 2\beta \Delta m^4\bigr) \biggr],
        \label{eq:4-probability-oscillation-nc}
\end{align}

\begin{figure}[h!]
    \centering
    \includegraphics[width=0.92\linewidth]{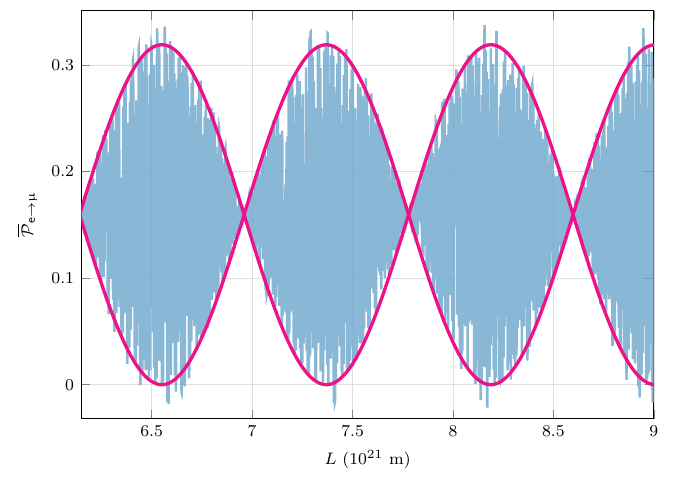}
        
    \caption{Plot of $\overline{\mathcal{P}}_{\mathsf{e}\to\upnu}$ for ${\theta = 0.3}$, ${\beta = 10^{-15}\, (\text{eV}^2 c^2)^{-1}}$, ${\Delta m^2 = 7.53\times 10^{-5} \ \text{eV}^2/c^4}$, ${E=1\ \text{MeV}}$ and ${m_1 = 0.1 \ \text{eV}/c^2}$.}\label{fig4-beat-mean-probability}
\end{figure}

In order to compare the neutrino flavor oscillation probability in the usual spacetime and in a spacetime with a presence of a minimal length, we calculate the mean value of equations \eqref{eq:4-probability-oscillation-commutative} and \eqref{eq:4-probability-oscillation-nc}, 
\begin{equation}\label{eq:4-mean-value-probability}
    \overline{\mathcal{P}}_{\mathsf{e} \to \upmu}
        \equiv \frac{1}{2} \bigl( \mathcal{P}_{\mathsf{e}\to\upmu}^{\text{C}} + \mathcal{P}_{\mathsf{e}\to\upmu}^{\text{NC}} \bigr).
\end{equation}
Equation \eqref{eq:4-mean-value-probability} is plotted in Figure \ref{fig4-beat-mean-probability}. At very long propagation distances ($\sim 10^{21}$ m), the comparison between the non-commutative and standard vacuum oscillation probabilities exhibits a characteristic beat pattern.

Assuming that Nature follows a non-commutative spacetime as described in Section \ref{sec2}, the oscillation of an astrophysical neutrino should follow the non-commutative probability in \eqref{eq:4-probability-oscillation-nc}. Comparing actual data with the commutative probability from simulations (using \eqref{eq:4-probability-oscillation-commutative}) would then reveal a beat pattern, an interference effect caused by the additional non-commutative phase $\delta_\text{NC}$. Conversely, if our spacetime is commutative, the comparison between actual data and simulations using \eqref{eq:4-probability-oscillation-commutative} would show no beat pattern. Therefore, up to current experimental limitations, we have a potential way of testing wether our universe is commutative or not.

In figure \ref{fig4-beat-mean-probability}, the blue curve indicates the numerical simulation of equation \eqref{eq:4-mean-value-probability}, while the pink curve indicates the analytical beat envelope, given by:
\begin{equation}\label{eq:4-beat-envelope}
    f_{\text{envelope}} = \frac{1}{2}\sin^2(2\theta)\bigg[1\pm \cos\bigg(\frac{L}{2\hbar E}\,\beta\Delta m^4 \bigg)\bigg].
\end{equation}
The envelope function sets the bounds of the oscillation probability. Therefore, values outside the pink curve indicate numerical imprecision and should be disconsidered. 

From \eqref{eq:4-beat-envelope}, we extract the beat pattern oscillation length
\begin{equation*}
    L_{\text{osc}} = \frac{4 \pi \hbar E}{ \beta\Delta m^4}.
\end{equation*}
Knowing the oscillation length, one can then determine the product $\beta \Delta m^4$ as
\begin{equation*}
    \beta \Delta m^4 = \frac{4 \pi \hbar E}{ L_{\text{osc}}}.
\end{equation*}


\section{Neutrino Oscillation in a Constant Magnetic Field}\label{sec4}
In this section, we first review the results obtained by \cite{Popov_2019} for neutrino oscillations in a constant magnetic field, and then introduce minimal length spacetime effects to oscillations in this scenario.

\subsection{In Ordinary Spacetime}
The Standard Model generalization for a Dirac neutrino diagonal magnetic moment \citep{Popov_2019} is
\begin{equation}\label{eq:6-neutrino-magnetic-moment}
    \mu_{ii} = \frac{3 e G_\text{F} m_i}{8\sqrt{2}\pi^2} \approx 3.2 \times10^{-19} \bigg( \frac{m_i}{1\ \text{eV}}\bigg)\mu_\text{B},
\end{equation}
where $e$ is the elementary charge, $G_\text{F}$ is the Fermi constant, and $\mu_\text{B}$ is the Bohr magneton.

To calculate the neutrino spin oscillation, let us begin by rewriting the two-flavor neutrino mixing equation using (\ref{eq:3-flavor-as-mass-mixing}) and the $2\times 2$ PMNS matrix in (\ref{eq:3-2-times-2-mixing-matrix}) with defined helicity (L or R). For electron ($\mathsf{e}$) and muon ($\upmu$) neutrinos, it reads as
\begin{subequations}
\begin{align}
    \label{eq:6-neutrino-eletron-state-stationary}
    |\upnu_\mathsf{e}^\text{L(R)}\rangle &= \cos\theta |\upnu_1^{\text{L(R)}}\rangle +  \sin\theta |\upnu_2^\text{L(R)}\rangle,
    \\
    \label{eq:6-neutrinon-muon-state-stationary}
    |\upnu_\upmu^\text{L(R)}\rangle &= - \sin\theta |\upnu_1^\text{L(R)}\rangle +  \cos\theta|\upnu_2^\text{L(R)}\rangle.
\end{align}
\end{subequations}
For an ultrarelativistic neutrino, its helicity states approximately coincide with its chiral states, i.e., $|\upnu_i^\text{L(R)}\rangle \approx |\upnu_i^{\text{ch}^-(\text{ch}^+)}\rangle$.

The effective Lagrangian density that describes a fermion with a magnetic moment $\mu$ with Pauli interaction is given by
\begin{equation}\label{eq:6-lagrangian-pauli-term}
    \mathcal{L}_{\text{mag}}
        = \bar{\Psi}\biggl(\mathrm{i}\hbar \gamma^\mu\partial_\mu - \frac{\mu}{2}\sigma^{\mu\nu}F_{\mu\nu} - m\mathbbm{1} \biggr)\Psi,
\end{equation}
where $F_{\mu\nu}$ is the background electromagnetic strength tensor. Picking an observer who sees only a static, purely magnetic field $\boldsymbol{B}$, and using the Pauli matrices relation $\sigma^{ij} = \frac{\mathrm{i}}{2}[\gamma^i, \gamma^j]$, the Lagrangian in \eqref{eq:6-lagrangian-pauli-term} takes the form:
\begin{equation}\label{eq:6-lagrangian-magnetic-field}
    \mathcal{L}_{\mathrm{mag}}
        = \bar{\Psi} \Bigl( \mathrm{i}\hbar \gamma^\mu\partial_\mu - \mu\,\boldsymbol{\Sigma}\cdot\boldsymbol{B} - m \Bigr) \Psi ,
\end{equation}
where $\boldsymbol{\Sigma} \equiv \gamma^5\gamma^0\boldsymbol{\gamma}$.

Let us now consider a massive neutrino propagating along the $z$-axis with momentum $\boldsymbol{p} = (0,0,p)$, in a constant homogeneous magnetic field $\boldsymbol{B} \equiv (\boldsymbol{B}_{\bot}, B_{\parallel})$, where $\boldsymbol{B}_{\bot}$ is the component on the $xy$-plane. The neutrino wave function in momentum space takes the form of a plane-wave solution to the Dirac equation obtained from the Lagrangian density \eqref{eq:6-lagrangian-magnetic-field}. For $s = \pm 1$, which are the eigenvalues of the spin operator, it follows that:
\begin{equation}\label{eq:6-dirac-equation-magnetic-field}
    \bigl( \gamma^\mu p_\mu - m_i - \mu_i\,\boldsymbol{\Sigma}\cdot\boldsymbol{B} \bigr) |\upnu_i^s(p)\rangle = 0,
\end{equation}
where $\mu_i$ is the neutrino magnetic moment. The $\mu_i\,\boldsymbol{\Sigma}\cdot\boldsymbol{B}$ term causes spin precession in a magnetic field, thereby producing neutrino spin-flavor oscillations. For the two-flavor case, we disregard $\mu_{ij}$ for $i \neq j$, which yields two decoupled equations for the states $|\upnu_i^s\rangle$. 

Equation (\ref{eq:6-dirac-equation-magnetic-field}) can be read as $\hat{H}|\upnu_i^s(p)\rangle = E|\upnu_i^s(p)\rangle $, where the Hamiltonian is given by
\begin{equation}\label{eq:6-hamiltonian}
    \hat{H}_i
        = \gamma^0 \, \boldsymbol{\gamma}\cdot\boldsymbol{p}
        + \mu_i\gamma^0 \, \boldsymbol{\Sigma}\cdot\boldsymbol{B}
        + m_i \gamma^0.
\end{equation}
Considering only positive energy solutions, the energy eigenvalues are
\begin{equation}\label{eq:6-energy-spectrum-with-magnetic-field}
    \begin{aligned}
    E_i^s &= \biggl( (m_i)^2 + p^2 + (\mu_i \boldsymbol{B})^2 + {}
    \\&\qquad{} + 2s \mu_i \sqrt{ (m_i \boldsymbol{B})^2 + (p B_{\bot})^2} \biggr)^{1/2},
    \end{aligned}
\end{equation}
where $\boldsymbol{B}^2 = B_{\parallel}^2 + \boldsymbol{B}_{\bot}^2$. For ultrarelativistic neutrinos, $p \gg m$ and $pc\gg \mu B$, thus the energy $E_i^s$ can be expanded as:
\begin{equation*}
    E_i^s \approx p + \frac{(m_i)^2}{2p} + \frac{(\mu_i\boldsymbol{B})^2}{2p} + s \mu_i B_{\bot} .
\end{equation*}
The neutrino magnetic moment $\mu_i$ is experimentally constrained to be very small, with current upper limits $\mu_\upnu \lesssim 10^{-11}\mu_\text{B}$ \citep{Borexino_2017}. For typical astrophysical magnetic fields, e.g., $B \sim 10^{13} - 10^{15}$ Gauss, as observed in pulsars \citep{Popov_2019}, the condition $\mu_i B \ll m_i$ holds even for neutrino masses in the eV range. Thus, $\mu_i^2\boldsymbol{B}^2/p$ is negligible compared to $m_i^2/p$, and the energy simplifies to:
\begin{equation}\label{eq:6-energy-spectrum-with-magnetic-field-ultrarelativistic-approx}
    E_i^s \approx E + \frac{(m_i)^2}{2E} + s \mu_i B_{\bot},
\end{equation}
where the ultrarelativistic approximation $E \approx p$ was used.

The spin operator $\hat{S}_i$ is chosen as in \cite{Ashutosh-2024}, namely
\begin{equation}\label{eq:6-spin-operator}
    \hat{S}_i
        = \frac{m_i}{\sqrt{ (m_i \boldsymbol{B})^2 + (p B_\bot)^2}} \biggl[ 
            \boldsymbol{\Sigma}\cdot\boldsymbol{B}
            - \frac{\mathrm{i}}{m_i} \, \gamma^0\gamma^5 \bigl(\boldsymbol{\Sigma}\times \boldsymbol{p}\bigr)\cdot \boldsymbol{B} \biggr] ,
\end{equation}
so that it commutes with the Hamiltonian in \eqref{eq:6-hamiltonian} and its eigenstates are stationary, providing the correct basis for describing neutrino spin evolution in a magnetic field. For neutrino stationary states, it is true that
\begin{equation*}
    \hat{S}_i^{} |\upnu_i^s\rangle = s|\upnu_i^s\rangle,
\end{equation*}
and
\begin{equation*}
    \langle \upnu_i^s|\upnu_j^{s'}\rangle = \delta_{ij}\delta_{s s'}.
\end{equation*}
From the spin operator in (\ref{eq:6-spin-operator}) we define the spin projection operator as:
\begin{equation}\label{eq:6-projection-operator}
    \hat{\Pi}_i^{\pm} = \frac{1\pm \hat{S}_i}{2},
\end{equation}
where it is trivial to see that the projection operator acts on neutrino mass stationary states as:
\begin{equation*}
     \langle \upnu_j^{s'}|\hat{\Pi}_i^{s}|\upnu_i^{s}\rangle = \delta_{ij}\delta_{s s'}.
\end{equation*}
Now, we want to analyze three possible neutrino oscillation in the presence of a magnetic field: flavor  $\upnu_\mathsf{e}^\text{L} \to \upnu_\upmu^\text{L}$, spin $\upnu_\mathsf{e}^\text{L} \to \upnu_\mathsf{e}^\text{R}$ and spin-flavor $\upnu_\mathsf{e}^\text{L} \to \upnu_\upmu^\text{R}$. Let us begin by expanding the neutrino helicity states over the neutrino stationary states\footnote{Note that helicity states $\nu_i^\text{L(R)}$ are eigenstates of the helicity operator $\frac{\boldsymbol{\sigma}\cdot \boldsymbol{p}}{|\boldsymbol{p}|}$, but they are not eigenstates of the Hamiltonian when $B\neq 0$.}:
\begin{subequations} \label{eq:6-expansion-neutrino-helicity-states}
\begin{align}
    \label{eq:6-expansion-neutrino-helicity-states-left}
    |\upnu_i^\text{L}(t)\rangle &= c_i^+  |\upnu_i^{+}(t)\rangle + c_i^-  |\upnu_i^{-}(t)\rangle,
    \\
    \label{eq:6-expansion-neutrino-helicity-states-right}
    |\upnu_i^\text{R}(t)\rangle &= d_i^+  |\upnu_i^{+}(t)\rangle + d_i^-  |\upnu_i^{-}(t)\rangle,
\end{align}
\end{subequations}

From \eqref{eq:6-expansion-neutrino-helicity-states}, it is possible to construct the time evolution of a relativistic state $|\upnu_\mathsf{e}^\text{L}(t)\rangle$,
\begin{equation}\label{eq:6-neutrino-eletron-time-evolution}
    \begin{aligned}
        |\upnu_\mathsf{e}^\text{L}(t)\rangle &= \cos\theta\Big(c_1^+ \mathrm{e}^{-\mathrm{i} E_1^+ t/\hbar}|\upnu_1^+\rangle + c_1^- \mathrm{e}^{-\mathrm{i} E_1^-t/\hbar}|\upnu_1^-\rangle \Big) + {} \\
        &\quad {} +\sin\theta \Big(c_2^+ \mathrm{e}^{-\mathrm{i} E_2^+ t/\hbar}|\upnu_2^+\rangle + c_2^- \mathrm{e}^{-\mathrm{i} E_2^-t/\hbar}|\upnu_2^-\rangle \Big).
    \end{aligned}
\end{equation}

The squared magnitudes of the coefficients $c_i^{\pm}$ and $d_i^{\pm}$, along with the cross-term $(d_i^\pm)^* c_i^\pm$, are determined by the matrix elements of the projection operator in \eqref{eq:6-projection-operator}; they can be determined from the initial conditions, due to coefficients $c_i^{\pm}$ and $d_i^{\pm}$ being independent of time. Thus, 
\begin{subequations} \label{eq:6-coefficient}
\begin{align}
    \label{eq:6-coefficient-c-2}
    |c_i^\pm|^2 &= \langle \upnu_i^\text{L}|\hat{\Pi}_i^\pm |\upnu_i^\text{L} \rangle,
    \\
    \label{eq:6-coefficient-d-2}
    |d_i^\pm|^2 &= \langle \upnu_i^\text{R}|\hat{\Pi}_i^\pm |\upnu_i^\text{R}\rangle,
    \\
    \label{eq:6-coefficient-d^*-c}
    (d_i^\pm)^*c_i^{\pm} &= \langle \upnu_i^\text{R}|\hat{\Pi}_i^\pm |\upnu_i^\text{L}\rangle.
\end{align}
\end{subequations}

The neutrino spinor structure can be normalized to unity in a box of volume $\ell_\text{norm}^3$, which yields to:
\begin{equation}\label{eq:6-spinor-structure-neutrino-normalized-by-ell}
    |\upnu_\text{L}\rangle_{\text{Dirac}} = \frac{1}{\sqrt{2}\ell_\text{norm}^{3/2}} \begin{pmatrix}
        1\\
        0\\
        -1\\
        0
    \end{pmatrix}, \quad
    |\upnu_\text{R}\rangle_{\text{Dirac}} = \frac{1}{\sqrt{2}\ell_\text{norm}^{3/2}}\begin{pmatrix}
        0\\
        1\\
        0\\
        1
    \end{pmatrix}.
\end{equation}
Substituting the spin operator (\ref{eq:6-spin-operator}), the spin projection operator (\ref{eq:6-projection-operator}) and the neutrino spinor structure (\ref{eq:6-spinor-structure-neutrino-normalized-by-ell}) into equations \eqref{eq:6-coefficient}, we write the quadratic coefficients as:
\begin{subequations}
\begin{align}
    \label{eq:6-coefficient-c2-expanded}
    |c_i^\pm|^2
        &= \frac{1}{2} \Biggl( 1\pm \frac{m_i B_{\parallel}}{\sqrt{(m_i B)^2 + (p B_{\bot})^2}} \Biggr),
    \\
    \label{eq:6-coefficient-d2-expanded}
    |d_i^\pm|^2
        &= \frac{1}{2} \Biggl(1\mp \frac{m_i B_{\parallel}}{\sqrt{(m_i B)^2 + (p B_{\bot})^2}} \Biggr),
    \\
    \label{eq:6-coefficient-c^*d-expanded}
    (d_i^\pm)^* c_i^\pm
        &= \mp \frac{1}{2} \frac{p\, (B_x -\mathrm{i}B_y)}{\sqrt{(m_i B)^2 + (p B_{\bot})^2}},
\end{align}
\end{subequations}

When the transversal magnetic field $\boldsymbol{B}_\bot$ is null, the helicity states are stationary and the coefficients are $|c_i^+|^2 = |d_i^-|^2=1$ and $|c_i^-|^2=|d_i^+|^2 =(d_i^\pm)^* c_i^\pm = 0$. In the ultrarelativist limit,
\begin{align*}
    |c_i^+|^2 &\approx |c_i^-|^2 \approx \frac{1}{4} ,
    \qquad (d_i^+)\,c_i^+ \approx - \frac{1}{4} ,
    \\
    |d_i^+|^2 &\approx |d_i^-|^2 \approx \frac{1}{4} ,
    \qquad
    (d_i^-)\,c_i^- \approx  \frac{1}{4} .
\end{align*}

Now, using the neutrino state time evolution in equation \eqref{eq:6-neutrino-eletron-time-evolution}, we can construct the neutrino oscillations in the presence of a magnetic field by substituting it into equation \eqref{eq:6-neutrino-eletron-time-evolution} and applying the result in equation \eqref{eq:6-expansion-neutrino-helicity-states-left}. For the neutrino flavor oscillation, it follows that
\begin{equation*}\label{eq:6-probability-spin-nua-e-crua}
    \begin{aligned}
        \mathcal{P}_{\mathsf{e},\text{L} \to \upmu,\text{L}}(t)
        &= \frac{1}{4}\sin^2(2\theta) \biggl| |c_2^+|^2\,\mathrm{e}^{-\mathrm{i}E_2^+ t/\hbar} + |c^-_2|^2\,\mathrm{e}^{-\mathrm{i}E_2^- t/\hbar} + {} \\
        &\qquad\qquad {}  -|c_1^+|^2\,\mathrm{e}^{-\mathrm{i}E_1^+ t/\hbar} - |c^-_1|^2\,\mathrm{e}^{-\mathrm{i}E_1^- t/\hbar} \biggr|^2,
    \end{aligned}
\end{equation*}
or yet
\begin{equation}\label{eq:6-flavor-oscillation-cos-sum-energy}
    \begin{aligned}
        \mathcal{P}_{\mathsf{e},\text{L} \to \upmu,\text{L}}
        &= \frac{1}{8}\sin^2(2\theta) \Biggl[ 2
        + \cos\biggl(\frac{L}{\hbar} \epsilon_{11}^{+-} \biggr)
        + \cos\biggl(\frac{L}{\hbar} \epsilon_{22}^{+-} \biggr) + {} \\
        &\qquad\qquad\qquad {}
        - \cos\biggl(\frac{L}{\hbar} \epsilon_{21}^{++} \biggr) 
        - \cos\biggl(\frac{L}{\hbar} \epsilon_{21}^{+-} \biggr) 
        + {} \\
        &\qquad\qquad\qquad {}
        - \cos\biggl(\frac{L}{\hbar} \epsilon_{21}^{-+} \biggr) 
        - \cos\biggl(\frac{L}{\hbar} \epsilon_{21}^{--} \biggr)
        \Biggr] ,
    \end{aligned}
\end{equation}
where the energy differences are
\begin{subequations} \label{eq:6-energy-differences-commutative}
\begin{alignat}{3}\label{eq:6-energy-differences-commutative-E1+-E1-}
    \epsilon_{11}^{+-} &\equiv
        E_1^+ - E_1^- &&= 2\mu_1 B_\perp ,
    \\ 
    \epsilon_{22}^{+-} &\equiv 
        E_2^+ - E_2^- &&= 2\mu_2 B_\perp ,
    \\
    \epsilon_{21}^{++} &\equiv 
        E_2^+ - E_1^+ &&= \frac{\Delta m^2}{2E} + (\mu_2 - \mu_1) \, B_\perp ,
    \\
    \epsilon_{21}^{+-} &\equiv 
        E_2^+ - E_1^- &&= \frac{\Delta m^2}{2E} + (\mu_2 + \mu_1) \, B_\perp ,
    \\
    \epsilon_{21}^{-+} &\equiv 
        E_2^- - E_1^+ &&= \frac{\Delta m^2}{2E} - (\mu_2 + \mu_1) \, B_\perp ,
    \\
    \epsilon_{21}^{--} &\equiv 
        E_2^- - E_1^- &&= \frac{\Delta m^2}{2E} - (\mu_2 - \mu_1) \, B_\perp .
    \label{eq:6-energy-differences-commutative-E2--E1-}
\end{alignat}
\end{subequations}

Analogously, the neutrino spin oscillation probability is
\begin{equation}\label{eq:6-spin-oscillation-probability-energy}
    \begin{aligned}
        \mathcal{P}_{\mathsf{e},\text{L} \to \mathsf{e},\text{R}}
        &= \frac{1}{2} \cos^4\theta \biggl[ 1 - \cos\biggl(\frac{L}{\hbar} \epsilon_{11}^{+-} \biggr) \biggr]
        + {} \\ 
        &\quad{} + \frac{1}{2}\sin^4\theta \biggl[ 1 - \cos\biggl(\frac{L}{\hbar} \epsilon_{22}^{+-} \biggr) \biggr]
        + {} \\
        &\quad{} + \frac{1}{8} \sin^2(2\theta) \biggl[ \cos\biggl(\frac{L}{\hbar} \epsilon_{21}^{++} \biggr)
            - \cos\biggl(\frac{L}{\hbar} \epsilon_{21}^{+-} \biggr) 
            + {} \\
            &\qquad\qquad\qquad{}
            - \cos\biggl(\frac{L}{\hbar} \epsilon_{21}^{-+} \biggr)
            + \cos\biggl(\frac{L}{\hbar} \epsilon_{21}^{--} \biggr)
        \biggr],
    \end{aligned}
\end{equation}
and the and spin-flavor oscillation probability is
\begin{equation}\label{eq:6-spin-flavor-oscillation-energy}
    \begin{aligned}
        \mathcal{P}_{\mathsf{e},\text{L} \to \upmu,\text{R}}
        &= \frac{1}{8} \sin^2(2\theta) \Biggl[ 2
        - \cos\biggl(\frac{L}{\hbar} \epsilon_{11}^{+-} \biggr) 
        - \cos\biggl(\frac{L}{\hbar} \epsilon_{22}^{+-} \biggr)
        + {} \\
        &\qquad\qquad\qquad{}
        + \cos\biggl(\frac{L}{\hbar} \epsilon_{21}^{++} \biggr)
        - \cos\biggl(\frac{L}{\hbar} \epsilon_{21}^{+-} \biggr)
        + {} \\
        &\qquad\qquad\qquad{}
        - \cos\biggl(\frac{L}{\hbar} \epsilon_{21}^{-+} \biggr)
        + \cos\biggl(\frac{L}{\hbar} \epsilon_{21}^{--} \biggr)
        \Biggr].
    \end{aligned}
\end{equation}

Note how \eqref{eq:6-flavor-oscillation-cos-sum-energy}, \eqref{eq:6-spin-oscillation-probability-energy} and \eqref{eq:6-spin-flavor-oscillation-energy} exhibit an explicit dependence on the neutrino masses $m_1$ and $m_2$ (rather than $\Delta m^2$), which follows from the definition of the magnetic moment in equation \eqref{eq:6-neutrino-magnetic-moment}. The same dependence is expected when we move to the non-commutative scenario in the next subsection.

\subsection{In a Minimal Length Spacetime}
To calculate the effects of the minimal length in the scenario where a constant magnetic field is involved, we apply the Samar--Tkachuk solution \eqref{eq:4-Samar-Tkachuk} into the Lagrangian density \eqref{eq:6-lagrangian-magnetic-field}, obtaining
\begin{equation}\label{eq:7-lagrangian-density-samar-tkachuk}
    \begin{aligned}
        \mathcal{L}^{\text{NC}}_{\text{mag}}
            &= \mathcal{L}_{\text{mag}}^{} + \frac{\mathrm{i}\hbar^3}{2} \beta \bigl[\bar{\Psi}\,\gamma^\mu\bigl(\square\,\partial_\mu \Psi\bigr) - \bigl(\square\,\partial_\mu \bar{\Psi}\bigr)\,\gamma^\mu\,\Psi \bigr] ,
    \end{aligned}
\end{equation}
where $\mathcal{L}_{\text{mag}}$ is given by \eqref{eq:6-lagrangian-magnetic-field} up to boundary terms. Comparing \eqref{eq:7-lagrangian-density-samar-tkachuk} with \eqref{eq:4-lagrangian-density-modified}, we see the introduction of the same correcting term in $\mathcal{O}(\beta)$.

The magnetic moment $\mu$ is kept unmodified, working as a coupling constant between the fermion (neutrino) and the background magnetic field. The modified Dirac equation in the presence of a magnetic field is, then,
\begin{equation}\label{eq:7-modified-dirac-equation-magnetic-field}
    \bigl[\mathrm{i}\hbar \gamma^\mu \bigl(1+\beta\hbar^2\Box\bigr)\partial_\mu - m - \mu\,\boldsymbol{\Sigma}\cdot \boldsymbol{B} \bigr]\Psi =0.
\end{equation}

Applying the plane wave ansatz \eqref{eq:4-plane-wave-ansatz} into the modified Dirac equation \eqref{eq:7-modified-dirac-equation-magnetic-field}, it yields
\begin{equation*}
    \biggl\{ \Bigl[ 1- \beta \bigl( E^2 - \boldsymbol{p}^2 \bigr) \Bigr] \bigl( \hat{\boldsymbol{\alpha}}\cdot\boldsymbol{p} - E\mathbbm{1} \bigr) + m\gamma^0 + \mu\gamma^0 \, \boldsymbol{\Sigma}\cdot\boldsymbol{B}\biggr\} \, \psi = 0 .
\end{equation*}
The energy eigenvalues are obtained when the determinant of the $4\times4$ matrix above vanishes, which will result in a sixth-degree polynomial in $E^2$.
The two light mass solutions for such polynomial are, up to first order in $\beta$,
\begin{equation}\label{eq:7-full-energy-solution}
    \begin{aligned}
        (E^s_{\text{NC}})^2 &= (E^s_{\text{C}})^2
        + 2\beta \Biggl[ (\mu\boldsymbol{B})^4 + m^4
        {} \\
        &\qquad {}
        + 2\mu^2 \bigl[ 3 (m\boldsymbol{B})^2 + (p B_\bot)^2 \bigr] +
        {} \\
        &\qquad {}
        + s \mu \, \frac{\bigl[4 (m\boldsymbol{B})^2 + 3 (p B_\bot)^2 \bigr] \bigl[(\mu \boldsymbol{B})^2 + m^2 \bigr]}{ \sqrt{(m\boldsymbol{B})^2 + (p B_\bot)^2}} \Biggr] ,
    \end{aligned}
\end{equation}
where $s=\pm1$ indicates the two possible light mass solutions, and $E^s_{\text{C}}$ is given by \eqref{eq:6-energy-spectrum-with-magnetic-field}.

Employing the same approximations described in Section \ref{sec5}, equation \eqref{eq:7-full-energy-solution} can be rewritten as:
\begin{equation}\label{eq:7-energy-nc-approx-value}
    E_\text{NC}^s
        \approx E_\text{C}^s
        + \beta \biggl[ \frac{m^4}{E} + 2(\mu B_\perp)^2E + 3s\mu B_\perp m^2 \biggr] .
\end{equation}

Assuming that $\mu_1 \approx \mu_2 \equiv \mu$, let us define the shorter notation
\begin{equation*}
    b \equiv \mu B ,
    \qquad
    b_\perp \equiv \mu B_\perp .
\end{equation*}
We can construct the energy differences $\mathcal{E}$ in the same way as in \eqref{eq:6-energy-differences-commutative}, but now with the non-commutative energies \eqref{eq:7-energy-nc-approx-value}. We find that
{\allowdisplaybreaks
\begin{align*}
    \mathcal{E}^{+-}_{11} &= 2b_\perp + 6\beta b_\perp (m_1)^2 , \\
    \mathcal{E}^{+-}_{22} &= 2b_\perp + 6\beta b_\perp (m_2)^2 ,\\
    \mathcal{E}^{++}_{21}&
        = \frac{1}{2E} \bigl( \Delta m^2 + 2\beta \Delta m^4 \bigr) ,\\
    \mathcal{E}^{+-}_{21} &= \frac{\Delta m^2}{2E} + 2 b_\perp + \beta \biggl[ \frac{\Delta m^4}{E} + 6 b_\perp \Delta m^2 \biggr] ,\\ 
    \mathcal{E}^{-+}_{21} &= \frac{\Delta m^2}{2E} - 2 b_\perp + \beta \biggl[ \frac{\Delta m^4}{E} - 6 b_\perp \Delta m^2 \biggr] , \\
    \mathcal{E}^{--}_{21}&= \frac{1}{2E} \bigl( \Delta m^2 + 2\beta \Delta m^4 \bigr) .
\end{align*}
}
Using the energy eigenstates \eqref{eq:7-energy-nc-approx-value} to include non-commutative effects, the flavor oscillation probability in a constant magnetic field becomes
\begin{equation}\label{eq:7-prob-flavor-NC-mu-delta}
    \begin{aligned}   
        \mathcal{P}^{\text{NC}}_{\mathsf{e},\text{L} \to \upmu,\text{L}}
        &= \frac{1}{4}\sin^2(2\theta) \Biggl\{ 1 + {}
        \\
        &+ \cos\biggl( 3\beta b_\bot \Delta m^2 \frac{L}{\hbar} \biggr) \cos\biggl( 2 b_\bot \bigl( 1 + 3\beta M^2 \bigr) \frac{L}{\hbar} \biggr)
        + {} \\
        &{} - \cos\biggl( \frac{1}{2E} \bigl( \Delta m^2 + 2\beta\Delta m^4 \bigr) \frac{L}{\hbar} \biggr)
        \times {} \\
        &\qquad\qquad{} \times \biggl[ 1 + \cos\biggl( 2 b_\perp (1 + 3\beta \Delta m^2) \frac{L}{\hbar}  \biggr) \biggr] \Biggr\} .
    \end{aligned}
\end{equation}
where $M^2 \equiv m_1^2 + m_2^2$.

For the spin oscillation, we have
\begin{equation}\label{eq:7-prob-spin-NC-mu-delta}
    \begin{aligned}
        \mathcal{P}_{\mathsf{e},\text{L} \to \mathsf{e},\text{R}}^\text{NC}
        &= \frac{1}{2}\cos^4\theta \biggl[
                1 - \cos\biggl( 2b_\perp \bigl(1+3\beta m_1^2 c^4\bigr) \frac{L}{\hbar} \biggr)
            \biggr]
        + {} \\
        &\quad{} + \frac{1}{2}\sin^4\theta \biggl[
                1 - \cos\biggl( 2b_\perp \bigl(1+3\beta m_2^2 c^4\bigr) \frac{L}{\hbar}\biggr) \biggr]
        + {} \\
        &\quad{} + \frac{1}{4} \sin^2(2\theta) \, \cos\biggl( \frac{1}{2E} \bigl( \Delta m^2 + 2\beta\Delta m^4 \bigr) \frac{L}{\hbar} \biggr)
        \times {} \\
        &\qquad\qquad{} \times \biggl[ 1- \cos\biggl( 2b_\perp \bigl(1 + 3\beta \Delta m^2 \bigr) \frac{L}{\hbar} \biggr) \biggr] ,
    \end{aligned}
\end{equation}
and finally, the spin-flavor oscillation probability is
\begin{equation}
    \begin{aligned}\label{eq:7-prob-spin-flavor-NC-mu-delta}
        \mathcal{P}_{\mathsf{e},\text{L} \to \upmu,\text{R}}^{\text{NC}}
        &= \frac{1}{4}\sin^2(2\theta) \Biggl\{ 1 + {} \\
        &{} - \cos\biggl( 3\beta b_\bot \Delta m^2 \frac{ L}{\hbar} \biggr) \cos\biggl( 2b_\bot \bigl( 1 + 3\beta M^2 \bigr) \frac{L}{\hbar} \biggr) 
        + {} \\
        &{} + \cos\biggl( \frac{1}{2E} \bigl( \Delta m^2 + 2\beta\Delta m^4 \bigr) \frac{L}{\hbar} \biggr)
        \times {} \\
        &\qquad\qquad {} \times
            \bigg[1 - \cos\biggl( 2b_\perp \bigl(1 + 3\beta \Delta m^2\bigr) \frac{L}{\hbar}\biggr) \biggr] \Biggr\} .
    \end{aligned}
\end{equation}
When $\beta=0$ the commutative probabilities are recovered.

\begin{figure}[h!]
    \centering
    \begin{minipage}{0.9\linewidth}
        \centering
        \includegraphics[width=0.92\linewidth]{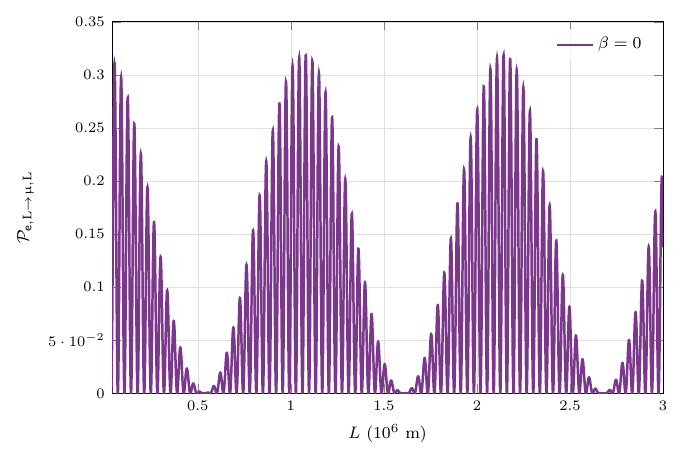}\label{fig:7-a-flavor}
    \end{minipage}
    \vspace{10pt}
    \begin{minipage}{0.9\linewidth}
        \centering
        \includegraphics[width=0.92\linewidth]{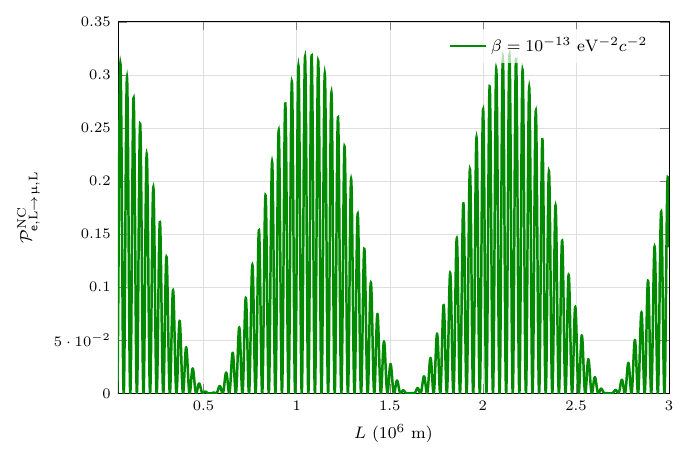}\label{fig:7-b-flavor}
    \end{minipage}
    \caption{\normalsize Probability of flavor oscillation for a left-handed electron neutrino converting to a left-handed muon neutrino. Parameters used: transversal magnetic field $B_\perp = 10^{12}\ \text{T}$, neutrino masses $m_1 = 0.1\ \text{eV}$ and $\Delta m^2 = 7.53 \times 10^{-5}\,\,\text{eV}^2/c^2$, magnetic moment $\mu=10^{-20}\, \mu_\text{B}$, mixing angle $\theta = 0.3$, and neutrino energy $E=1\ \text{MeV}$. The purple curve (upper panel) corresponds to the commutative scenario ($\beta=0$), while the green curve (lower panel) represents a non-commutative spacetime with $\beta=10^{-13}\ \text{eV}^{-2} c^{-2}$.}\label{fig:6-oscillation-flavor}
\end{figure}

Figure \ref{fig:6-oscillation-flavor} shows the flavor oscillation probability for neutrinos propagating in a commutative spacetime (upper panel), computed from \eqref{eq:6-flavor-oscillation-cos-sum-energy}, and in a non-commutative spacetime with $\beta = 10^{-13}\ \text{eV}^{-2} c^{-2}$ (lower panel), given by \eqref{eq:7-prob-flavor-NC-mu-delta}.

\begin{figure}[h!]
        \centering
        \includegraphics[width=0.92\linewidth]{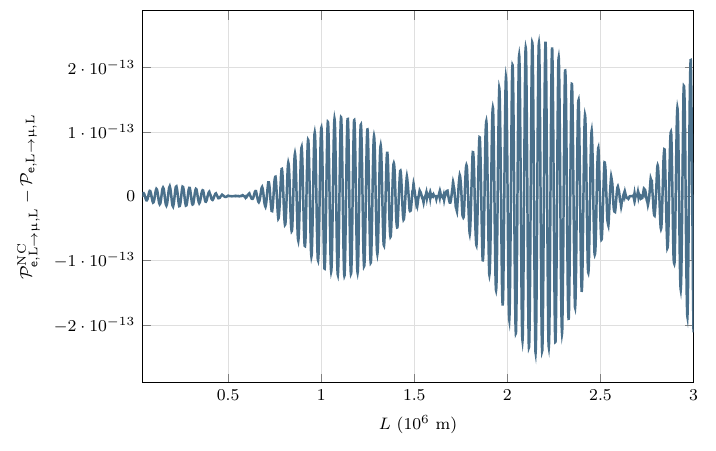}
        \caption{Difference between the non-commutative flavor oscillation probability ($\beta = 10^{-13}\, \text{eV}^{-2}\text{c}^{-2}$) and and the commutative probability ($\beta=0$), derived from the curves shown in \ref{fig:6-oscillation-flavor}.}\label{fig:6-flavor-difference-of-probabilities}
\end{figure}

Although the two oscillation probabilities in Figure \ref{fig:6-oscillation-flavor} appear similar at first glance, subtracting them yields the curve shown in Figure \ref{fig:6-flavor-difference-of-probabilities}. This result indicates that, for a propagation distance of $10^3\ \text{km}$, non-commutativity of spacetime, characterized by the parameter $\beta = 10^{-13}\ \text{eV}^{-2} c^{-2}$ and a minimal length $\ell_{\text{min}} \sim 10^{-22}\ \text{m}$, induces a deviation of approximately $10^{-13}$ in the oscillation probability relative to the standard commutative case.

\begin{figure}[h!]
    \centering
    \begin{minipage}{0.9\linewidth}
        \centering
        \includegraphics[width=0.92\linewidth]{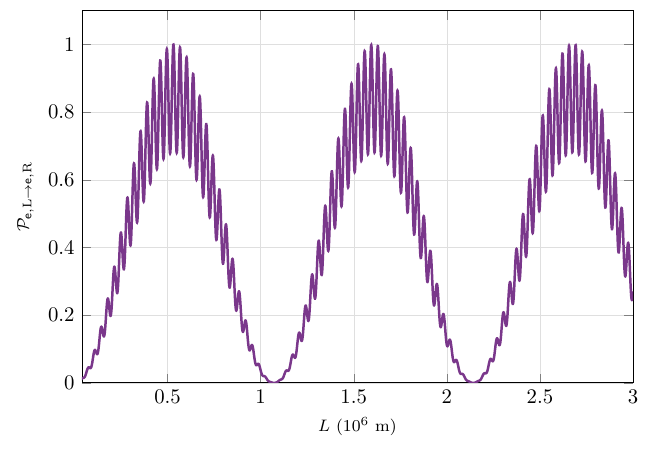}\label{fig:7-a-spin}
    \end{minipage}
    \vspace{10pt}
    \begin{minipage}{0.9\linewidth}
        \centering
        \includegraphics[width=0.92\linewidth]{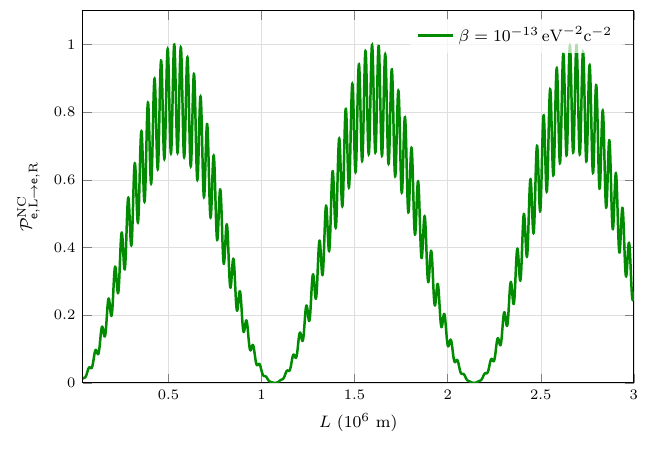}\label{fig:7-b-spin}
    \end{minipage}
    \caption{Probability of spin oscillation of a left-handed electron neutrino to a right-handed electron neutrino. For transverse magnetic field $B_\perp=10^{12}$ T, $m_1 = 0.1$ eV, $\Delta m^2 = 7.53 \times 10^{-5}\,\,\text{eV}^2/c^2$, $\mu=10^{-20}\mu_\text{B}$, $\theta = 0.3$ and $E=1$ MeV. The purple curve (upper panel) indicates the commutative scenario, when $\beta=0$, and the green one (lower panel) indicates a non-commutative spacetime with $\beta=10^{-13}\, \text{eV}^{-2}\text{c}^{-2}$.}\label{fig:6-oscillation-spin}
\end{figure}

Following the same approach, the plots for the neutrino spin oscillation in the commutative spacetime given by \eqref{eq:6-spin-oscillation-probability-energy}, and in the non-commutative spacetime given by \eqref{eq:7-prob-spin-NC-mu-delta}, are shown in figure \ref{fig:6-oscillation-spin}.

\begin{figure}[h!]
        \centering
        \includegraphics[width=0.92\linewidth]{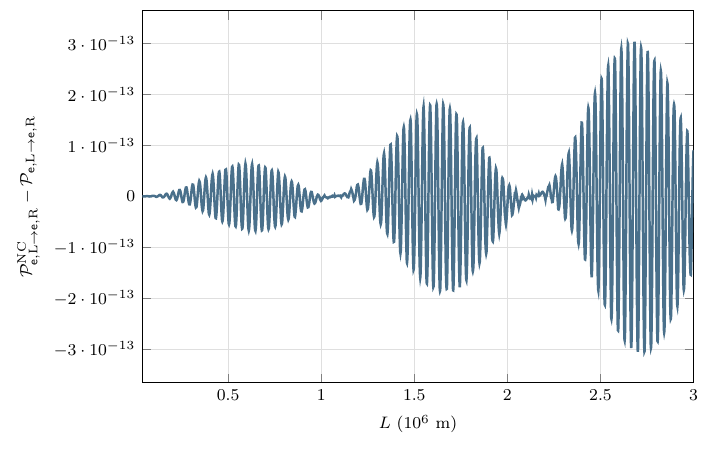}
        \caption{Plot of the difference of non-commutative probability ($\beta = 10^{-13}\, \text{eV}^{-2}\text{c}^{-2}$) and commutative probability ($\beta=0$) for neutrino spin oscillation. Difference of the two curves in figure \ref{fig:6-oscillation-spin}.}\label{fig:6-spin-difference-of-probabilities-spin}
\end{figure}

The difference between the commutative and the non-commutative probabilities for the neutrino spin oscillation is shown in the plot of figure \ref{fig:6-spin-difference-of-probabilities-spin}. Again, we note a deviation of order $10^{-13}$.

\begin{figure}[h!]
    \centering
    \begin{minipage}{0.9\linewidth}
        \centering
        \includegraphics[width=0.92\linewidth]{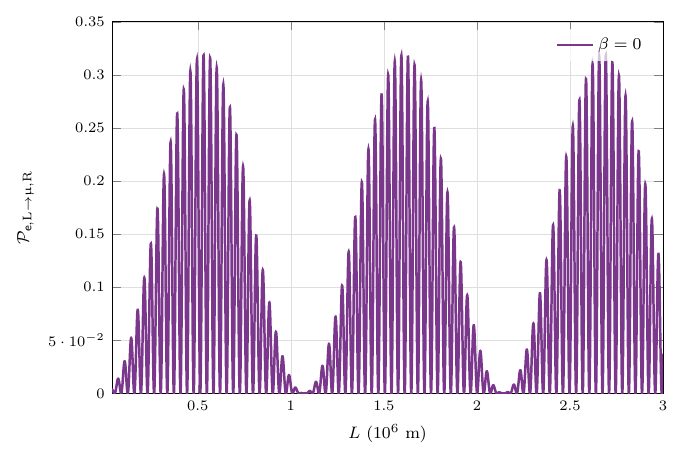}\label{fig:7-a-spin-flavor}
    \end{minipage}
    \vspace{10pt}
    \begin{minipage}{0.9\linewidth}
        \centering
        \includegraphics[width=0.92\linewidth]{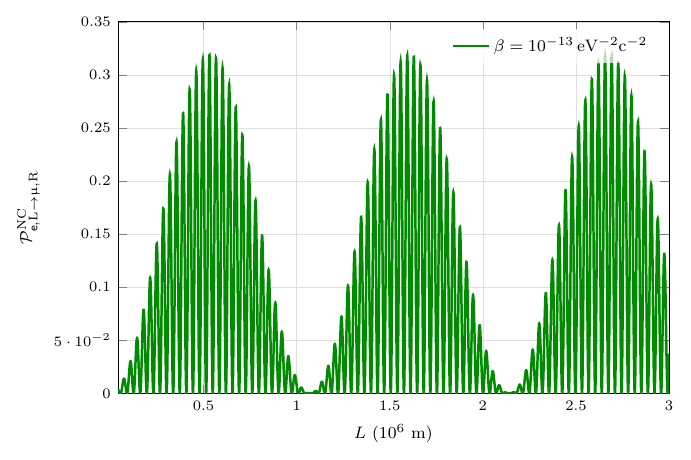}\label{fig:7-b-spin-flavor}
    \end{minipage}
    \caption{Probability of spin-flavor oscillation of a left-handed electron neutrino to a right-handed muon neutrino. For transversal magnetic field $B_\perp=10^{12}$ T, $m_1 = 0.1$ eV, $\Delta m^2 = 7.53 \times 10^{-5}\,\,\text{eV}^2/c^2$, $\mu=10^{-20}\mu_\text{B}$, $\theta = 0.3$ and $E=1$ MeV. The purple curve (upper panel) indicates the commutative scenario, when $\beta=0$, and the green one (lower panel) indicates a non-commutative spacetime with $\beta=10^{-13}\, \text{eV}^{-2}\text{c}^{-2}$.}\label{fig:6-oscillation-spin-flavor}
\end{figure}

\begin{figure}[h!]
        \centering
        \includegraphics[width=0.92\linewidth]{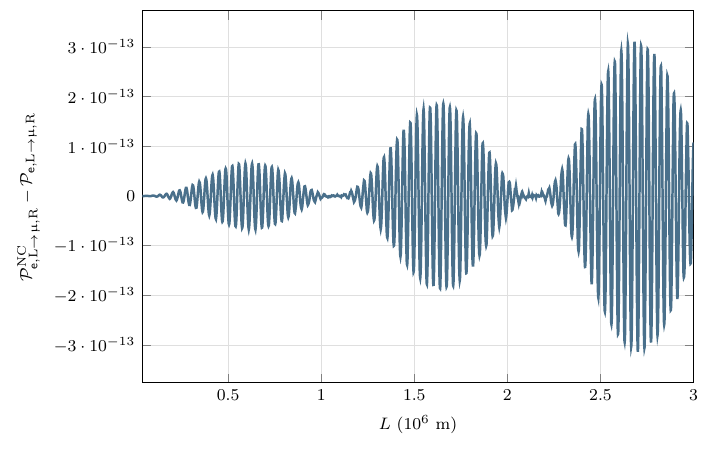}
        \caption{Plot of the difference of non-commutative probability ($\beta = 10^{-13}\, \text{eV}^{-2}\text{c}^{-2}$) and commutative probability ($\beta=0$) for neutrino spin oscillation. Difference of the two curves in figure \ref{fig:6-oscillation-spin-flavor}.}\label{fig:6-spin-difference-of-probabilities-spin-flavor}
\end{figure}

The same behavior is observed for the neutrino spin-flavor oscillation. Figure \ref{fig:6-oscillation-spin-flavor} shows the plot of equations \eqref{eq:6-spin-flavor-oscillation-energy} and \eqref{eq:7-prob-spin-flavor-NC-mu-delta}, while figure \ref{fig:6-spin-difference-of-probabilities-spin-flavor} shows the difference of the oscillation probabilities of the two scenarios. 

\section{Conclusion}\label{sec5}

This paper investigated neutrino flavor oscillations in vacuum and flavor, spin, and spin-flavor oscillations in strong background magnetic fields within a minimal length spacetime.

In vacuum, following \cite{moayedi-2011}, we derived an effective neutrino mass via the Samar--Tkachuk solution to the Dirac Lagrangian. Incorporating it into the standard oscillation probability yields a beat pattern at long distances when comparing commutative and non-commutative cases, allowing an estimation of $\beta \Delta m^4$.

For oscillations in the presence of a background magnetic field, we obtained a new energy solution from the fermion Lagrangian with magnetic moment $\mu$ under non-commutativity. Applying it to oscillation probabilities, with minimal length effects in the neutrino mass and $\mu$ as a coupling constant, we found a deviation of order $10^{-13}$ between standard and non-commutative probabilities for flavor, spin, and spin-flavor oscillations.


\bibliography{Bibliography}%


\section*{Acknowledgements}

This work was based on LNC master's thesis developed in the Universidade Federal do Rio Grande do Sul (UFRGS). LNC
would like to thanks UFRGS and Capes for financial support.

TOF is grateful for the hospitality of Perimeter Institute where part of this work was carried out. Research at Perimeter Institute is supported in part by the Government of Canada through the Department of Innovation, Science and Economic Development and by the Province of Ontario through the Ministry of Colleges and Universities.

\end{document}